\documentclass{caosp302}
\usepackage{graphicx}
\usepackage{natbib,openbib}
\usepackage{txfonts}
\articleNo{123}
\pubyear{2005}
\volume{35}
\volnumber{3}
\firstpage{1}
\received{November 21, 2007}
\accepted{November 30, 2007}

\begin{document}
\newcommand{\scrbox}[1]{\ensuremath{{\mbox{\scriptsize #1}}}}

\newcommand{\teff}{{\ensuremath{T_{\scrbox{eff}}}}}
\newcommand{\msol}{\ensuremath{\,M_{\odot}}}
\newcommand{\Mstar}{\ensuremath{\,M_{*}}}
\newcommand{\Mmixed}{\ensuremath{\Delta M_{\rm m}}}
\newcommand{\Dturb}{\ensuremath{D_{\scrbox{T}}}}
\newcommand{\Kelvin}{\,\mbox{K}}
\newcommand{\MS}{main--sequence}
\newcommand{\gr}{\ensuremath{g_{\scrbox{rad}}}}
\newcommand{\tbcz}{{\ensuremath{T_{\scrbox{bcz}}}}}
\newcommand{\mbcz}{{\ensuremath{M_{\scrbox{bcz}}}}}
\newcommand{\DM}{\ensuremath{ \log \Delta M/M_{*}}}
\newcommand{\DMsol}{\ensuremath{ \log \Delta M/M_{\odot}}}

\renewcommand{\H}{\mbox{H}}
\newcommand{\He}{\mbox{He}}
\newcommand{\Be}{\mbox{Be}}
\newcommand{\B}{\mbox{B}}
\newcommand{\Fe}{\mbox{Fe}}
\newcommand{\Mn}{\mbox{Mn}}
\newcommand{\Mg}{\mbox{Mg}}
\newcommand{\Si}{\mbox{Si}}
\newcommand{\Cr}{\mbox{Cr}}
\newcommand{\Ca}{\mbox{Ca}}
\newcommand{\Ni}{\mbox{Ni}}
\newcommand{\Ne}{\mbox{Ne}}
\newcommand{\Ti}{\mbox{Ti}}
\newcommand{\F}{\mbox{F}}
\newcommand{\Ox}{\mbox{O}}
\newcommand{\K}{\mbox{K}}
\newcommand{\Li}{\mbox{Li}}

\newcommand{\mix}{_{\rm{mix}}}
\newcommand{\T}{_{\rm{T}}}
\newcommand{\ad}{_{\rm{ad}}}
\newcommand{\rad}{_{\rm{rad}}}
\newcommand{\Le}{_{\rm{L}}}
\newcommand{\SC}{_{\rm{SC}}}
%
\hauthor{G. Michaud and J. Richer}

\title{Horizontal Branch stars as AmFm/HgMn stars}


%
\author{
        G.\,Michaud \inst{1,} \inst{2,} 
      \and 
        J.\,Richer \inst{1}   
             }

%
\institute{
           D\'epartement de Physique, Universit\'e de Montr\'eal, C. P. 6128,
       Montr\'eal, PQ, H3C~3J7, CANADA \email{michaudg@astro.umontreal.ca; jacques.richer@umontreal.ca}
         \and 
         LUTH, Observatoire de Paris, CNRS, Universit\'e Paris Diderot, 
     5 Place Jules Janssen,
     92190 Meudon, FRANCE
          }

\date{March 8, 2003}

\maketitle

\begin{abstract} Recent observations and models for horizontal branch stars are briefly described and compared to models 
for AmFm stars.  The limitations of those models are emphasized by a 
comparison to observations and models for HgMn stars.
\end{abstract}

%
\section{Introduction}
\label{sec:intr}
At around 16000\,K, the \teff{} where the horizontal branch (HB) crosses the main--se\-quen\-ce, HB stars have the same \teff{} and $\log g$ as HgMn stars.
Both have very shallow surface convection zones and they appear to have some abundance anomalies in common \citep{SargentSe67,SargentSe68}. At slightly cooler \teff, HB stars have a smaller $\log g$ and their convection zones become as deep or even deeper  than those of AmFm stars.
On the \MS{}, stellar evolution  models  assuming a mixed outer zone have been relatively successful at explaining 
AmFm stars \citep{RicherMiTu2000,MichaudRiRi2005}.  Detailed models for HgMn stars are more complex and have involved NLTE calculations
of individual species (see \citealt{Michaud81} for a review but also \citealt{MichaudReCh74} and \citealt{ProffittBrLeetal99}).  

We first very briefly (\S\,\ref{sec:calculations}) describe stellar evolution calculations done with all effects of atomic diffusion including radiative accelerations, \gr. We then show applications to Pop II stars (\S \ref{sec:PopII}).  We will briefly describe recent observations of abundance anomalies in HB stars and  what fraction of 
them can be explained by stellar evolution models with a mixed outer zone (\S \ref{sec:Anomalies}), a model similar to that used for AmFm stars.  After a few examples of results  for AmFm and HgMn stars (\S \ref{sec:PopI}), the role of chemical differentiation  in the atmosphere of HB stars is discussed in \S \ref{sec:Conclusion} as well as the potential role of rotation in reducing anomalies, thus linking HB and HgMn stars.

\section{Stellar evolution with radiative accelerations}
\label{sec:calculations}
The particle transport equations are introduced into a stellar evolution code which was descibed in \citet{VandenBerg85}, \citet{ProffittMi91} and \citet{Proffitt94}.
For each species one adds a force equation (eq. [18.1] of \citealt{Burgers69}) and a heat equation (eq. [18.2] of Burgers). There are consequently two coupled differential equations for each of the 28 included species.  Similar equations are written for electrons. 
 It is generally assumed that each atomic species can be treated  locally as being in an average state of ionization.
One  needs to know $Z_i$, an appropriate mean of the number of lost electrons.

The dominant term for each species contains $\gr(A) - g$ as a factor, where $\gr(A)$ is an appropriate average 
of the radiative acceleration over the states of ionization of element $A$.  Over most of the stellar interior and for most of the evolution, it dominates transport even though the electric field, ``diffusion'' and thermal diffusion terms are also included in the calculations and are important for part of the evolution in some stars. 

Rosseland opacities, mean ionic charges, and mean radiative forces are calculated
using the same interpolation method, based on the principle of corresponding
states described in  \S\,2.2 of \citet{RogersIg92a}; see in particular their equations (4) to (6).   Interpolation
weights are determined for a subset of the data grid, and used to interpolate locally all
these variables.

In first approximation, evaluating  $\gr(A)$ amounts  to
calculating the fraction of the momentum flux that each element absorbs from the photon flux.  In
stellar interiors, it takes the form:
\begin{equation}
\gr(A) = {{L_r^{\mbox{\scriptsize rad}}}\over{4\pi r^2 c}}
   \frac{\kappa_{R}}{X_A} \int_0^\infty
   \frac{\kappa_u(A)}{\kappa_u(\mbox{total})} { \mathcal{P}}(u)du
 \label{eq:grad_simple_def}
\end{equation}   
where most symbols have their usual meaning.  The quantities  $\kappa_u(\mbox{total})$ and $\kappa_u(A)$
are respectively the total opacity and the contribution of element $A$ to the total opacity at frequency $u$, with  $u$ and ${\mathcal{P}}(u)$  given by:
\begin{equation}
u={{h\nu}\over{kT}}
\label{eq:u}
\end{equation}
and
\begin{equation}
{\mathcal{P}} \left({u}\right)={{15}\over{4\pi^4}}  {u^{4}{{e^{u}}\over{{\left({e^{u}-1}\right)}^{2}}}}.
\label{eq:pu}
\end{equation}
The calculations of \gr($A$) involve carrying out the integration over the 10$^4$ $u$ 
values for each atomic species, $A$. 
It implies using a 1.5 Gigabyte spectrum data base from OPAL. The integrations must be repeated for each atomic species, at each mesh point (typically 2000 in our models) and at each time step (typically 10$^4$ up
to the giant branch). The Rosseland average opacity is also continuously recalculated making these calculations fully self consistent with all composition changes. 

\section{Pop II and HB stars}
\label{sec:PopII}
Evolutionary calculations have been carried out for a large number of masses and metallicities \citep{RichardMiRi2002,RichardMiRietal2002}.  These have in particular been used to determine the age of M92, one of the oldest globular clusters, to be 13.5 Gyr,  and to constrain the Li abundance to be expected from stellar evolution in the oldest stars.  Both were compared to observations of WMAP \citep{RichardMiRi2005,VandenBergRiMietal2002}.

\subsection{From the Main Sequence to the tip of the Giant Branch}
\label{sec:MainSequence}
Here, we follow a 0.8\msol{} model with $Z = 10^{-4}$ from the zero age \MS{} to the middle of the HB.  Its metallicity allows a comparison to observations in the globular cluster M15.
The evolution is carried out without the adjustment of any parameter except in so far as the mixing length was determined to fit the solar model \citep{TurcotteRiMietal98}.

During \MS{} evolution, concentration variations develop throughout the star.  At turnoff, a 0.8\msol{} star 
has $\teff \sim 6500$\,K and an age $\sim 11.5$\,Gyr \citep{RichardMiRietal2002}.  The concentration variations 
 caused by atomic diffusion extend over 30\%\,of the radius at the level of a factor of 1.5 in contrast and over 60\%\, of the radius at the level 1.2 in contrast.  There are  overabundances  of all atomic species from Na to Ni in the atmospheric regions but overabundance factors vary among the species.  Some of those abundance anomalies may have been seen in M92 by \citet{KingStBoetal98} but at the limit of detection so that  this result requires confirmation since the signal to noise ratio was not quite satisfactory.  Metals pushed from the deeper interior also accumulate within the star 
 where $\gr{} - g \sim 0$.  This occurs
progressively deeper in, as one considers species with larger atomic numbers  from S to Ni, since a given electronic configuration occurs progressively deeper in the star as the atomic number increases. 

As the evolution proceeds on the subgiant and giant branches, dredge up occurs and mixes some 58\,\%  of the star.
The abundance anomalies are largely but not completely eliminated.  At the He flash, there remains a 0.04 dex difference
in metal abundance concentration between the surface and the core and  a 0.003 \msol{} difference in helium core mass between a model calculated with and one without diffusion.
  
\subsection{Horizontal Branch Stars}
\label{sec:HB}
The transition to the HB was carried out  as described in \citet{MichaudRiRi2007}:  
a fraction of the envelope mass was removed and the star was reconverged on the HB
following approximately the procedure used by \citet{Sweigart87}.
On the HB there are effects of diffusion in three regions of  the star: just outside the He burning core, just below the H burning shell and to 
produce surface abundance anomalies.  The first two are discussed in \S\,\ref{sec:Structural} and the last one in 
\S\,\ref{sec:Anomalies}.

\subsubsection{Structural effects of diffusion}
\label{sec:Structural}
In canonical stellar evolution it has been standard practice since the suggestion of \citet{Paczynski70} 
to assume that the He burning core is extended by overshooting or penetration in order to maintain convective neutrality at the boundary. Since the opacity of carbon rich material is larger than that of He rich material, the radiative gradient increases in the core as He burning transforms He into C.  If the core boundary is stationary, 
strong superadiabaticity develops at the boundary.  It has been customary to assume that this causes overshooting.  This process is however poorly understood (see \citealt{Sweigart94}) and its efficiency is essentially unknown.  In stellar evolution with atomic diffusion, this superadiabaticity does not develop.  Instead atomic diffusion,
even in the absence of overshooting, is sufficient to cause core expansion and maintain convective neutrality at the boundary (see Fig. [10] of \citealt{MichaudRiRi2007}).  This does not prove that overshooting does not exist, but
 it is not necessary to ascribe to it a significant mixing efficiency since the transport
it was assumed to carry can be done by atomic diffusion when it is properly included.

The second structural effect of atomic diffusion is related to H burning.  Atomic diffusion  of the H which is burning in the shell causes an extension
inward of this shell. Hydrogen makes contact with C that was synthesized during the
He flash. It leads to an increase in H burning and so of luminosity but only of $\sim 0.01$\,dex.

Those structural effects of diffusion are caused by \emph{atomic} diffusion.  They are not influenced by the turbulent transport 
that was introduced in the  outer envelope as described in \S \ref{sec:Anomalies}.

\begin{figure}[t!]
\centerline{
\includegraphics[width=0.5\hsize]{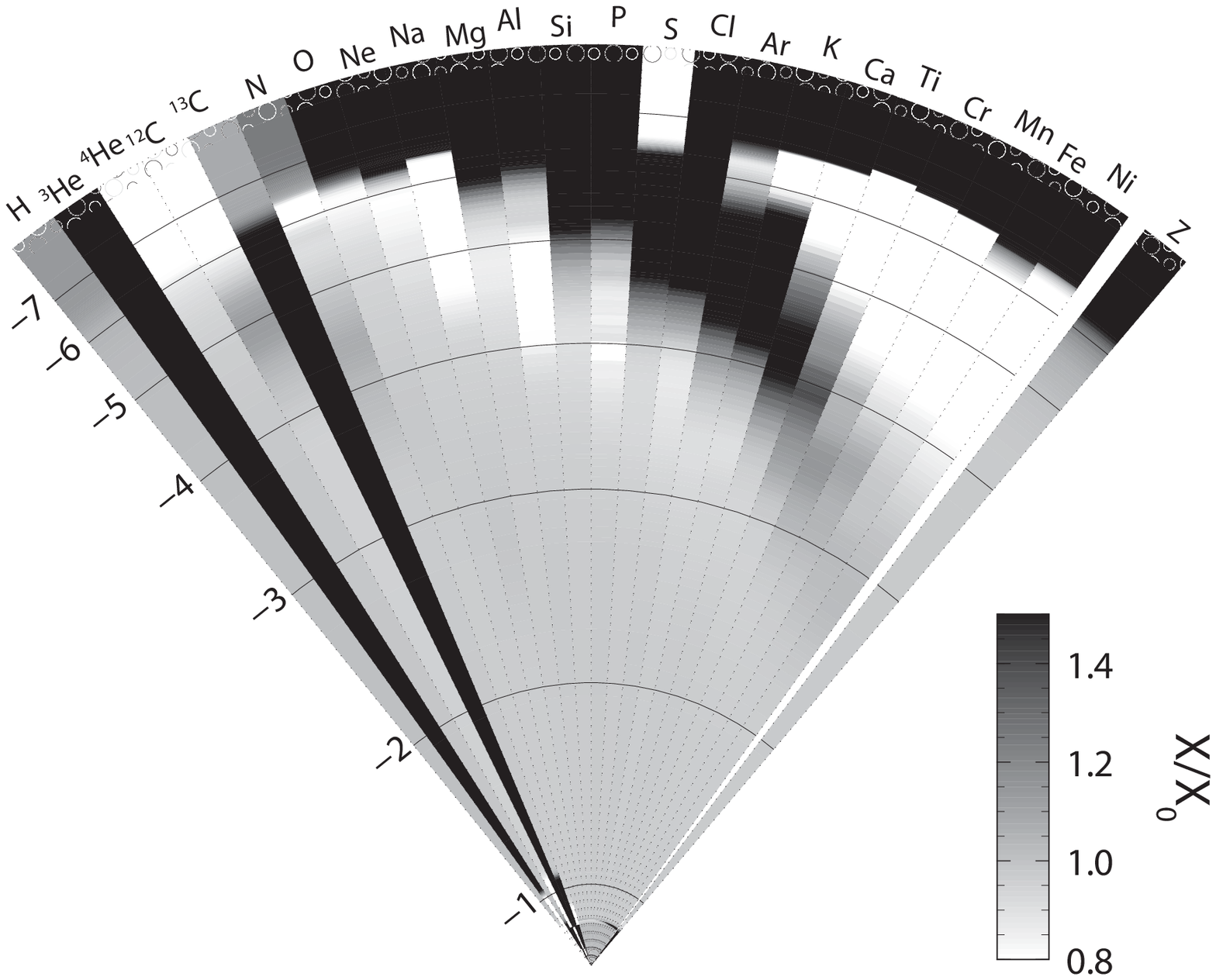}
\includegraphics[width=0.5\hsize]{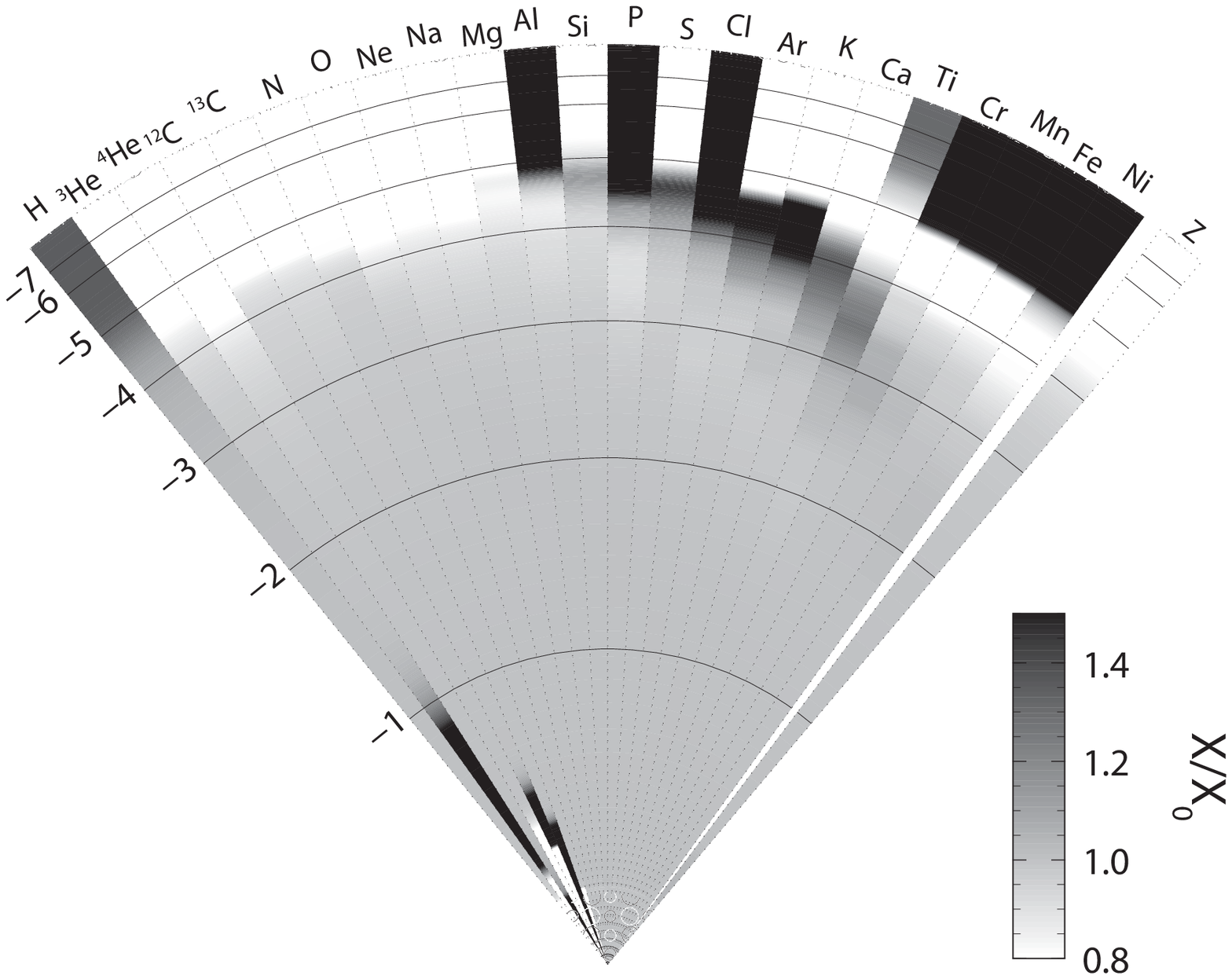}}
\caption{
Concentration variations for all calculated species, \emph{letft panel}, in a 0.61 \msol{} model after 30 Myr on the HB
($\teff \sim 12 400$\,K) and, \emph{right panel}, in a 2.0 \msol{} Pop I star ($\teff \sim 8005$\,K) after 616 Myr on the \MS{} as appropriate for a Hyades star.  The gray scale is adjusted so that an overabundance by a factor of 1.5 or more appears black in both cases while an underabundance by a factor of 0.8 or less appears white. The radial scale is linear in $r$. Horizontal lines indicate  the mass of the spherical shell  outside a certain radius ($\Delta m$) labeled by $\log (\Delta m/\Mstar)$.  The HB model is more concentrated since there
is about $\log (\Delta m/\Mstar) = -2$ outside of the fractional radius where there is $\log (\Delta m/\Mstar) = -1$ in the \MS{} model.  This leads to larger effects of diffusion on the HB.
The outer 50\% by radius is affected by diffusion in the HB model $(\log (\Delta m/\Mstar) = -3)$ while it is the outer 25 \% by radius in the Pop I model $(\log (\Delta m/\Mstar) = -3)$ even if 20 times longer was available to the Pop I model.}
\label{fig:EventailHB}
\end{figure}
\begin{figure}[t!]
\centerline{
\includegraphics[width=0.45\hsize]{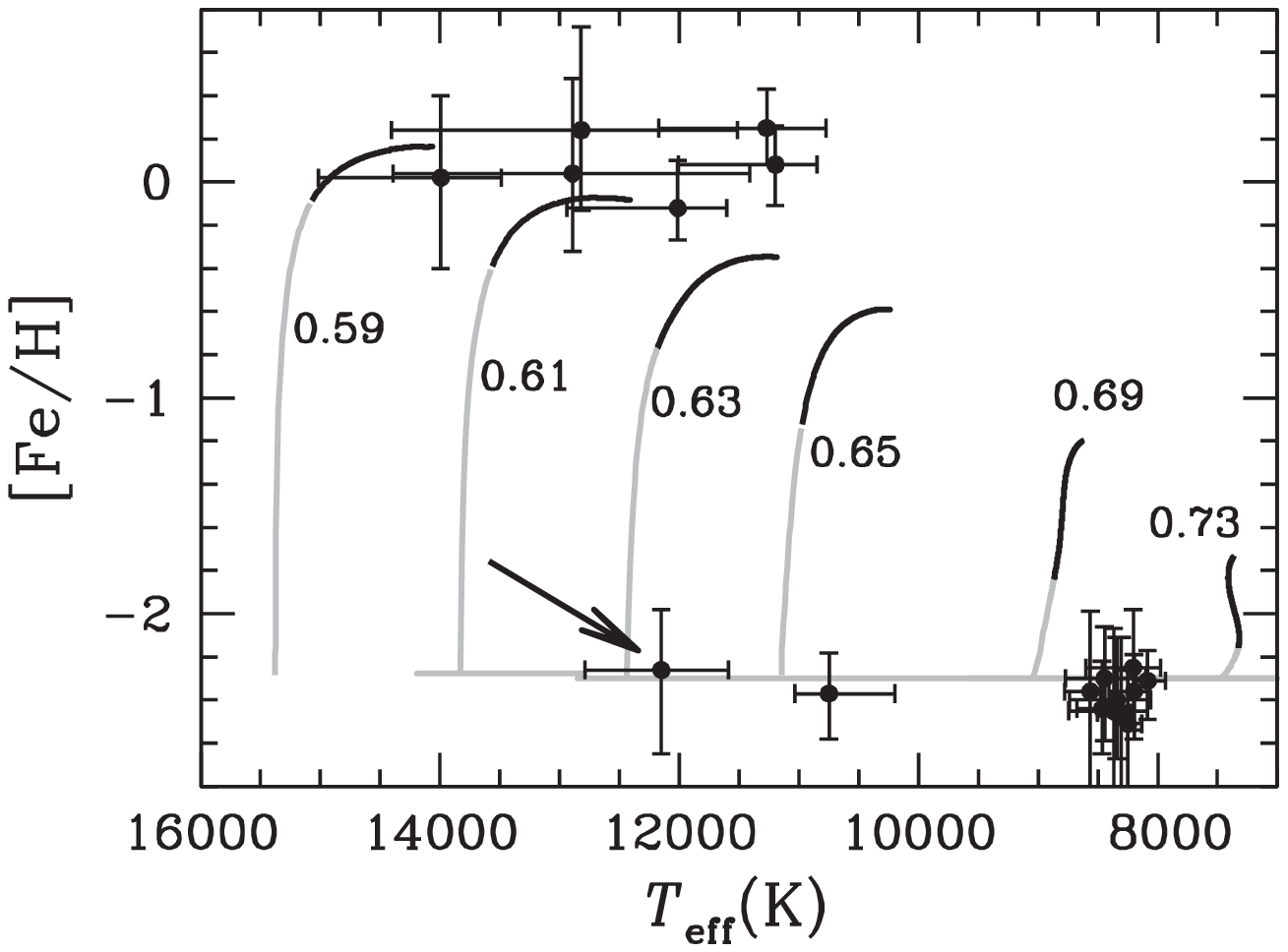}
\hspace{\fill}
\includegraphics[width=0.5\hsize]{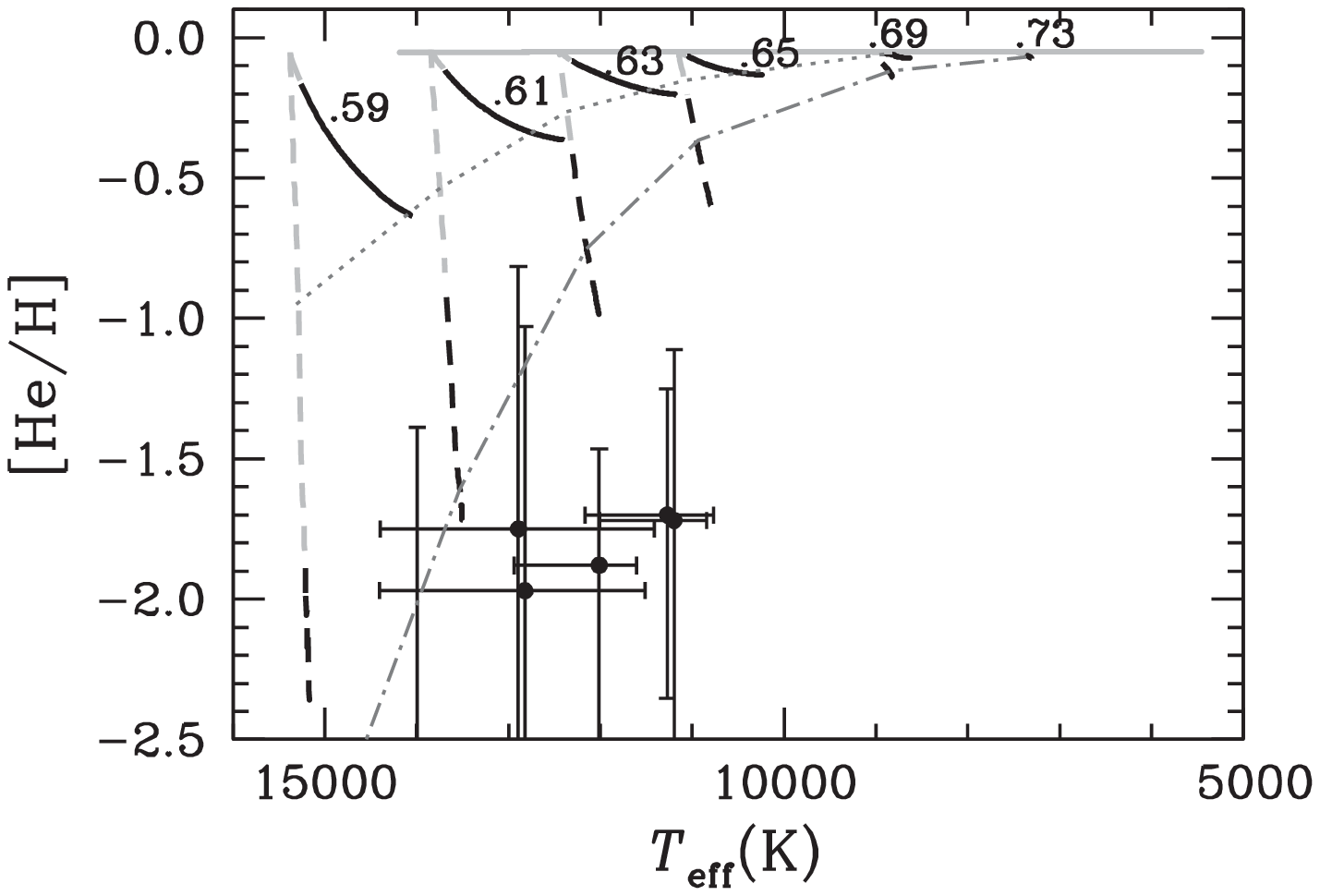}}
\caption{
\emph{Left panel}, concentration of surface [Fe/H] expected in HB models compared to observations of \citet{Behr2003}
for M15.
The continuous dark lines cover the interval from 5 to 30 Myr after zero age HB (ZAHB) for a number of models with a turbulence mixing the outer stellar region with $T<100\,000$\,K. The mass of each model is identified on the figure in \msol.  The star marked with an arrow has $V \sin i \sim 16$ km s$^{-1}$.
All other stars with $\teff > 11000$\,K have $V \sin i <$\,8\,km\,s$^{-1}$.  
\emph{Right panel}, concentration of [He/H] expected in two series of  HB models compared to observations of Behr for M15.  The continuous lines are defined similarly as for [Fe/H] and were calculated with a turbulence that mixes the outer stellar region with $T<100\,000$\,K.  The dashed lines were calculated with a turbulence that mixes the outer stellar region with $T<30\,000$\,K. The dotted line links models at 1 Myr after ZAHB 
while the dashed-dotted line links models at 
5 Myr.  One notes that the settling of He is some 30 $\times$ faster if the star is stable up to 30\,000\,K.}
\label{fig:Fe}
\end{figure}
\subsubsection{Surface abundance anomalies}
\label{sec:Anomalies}
 Since HB stars are Pop II stars that just left the  giant branch of globular clusters, they are all expected to have the same 
concentration of metals, at least of those heavier than Al \citep{GrattonSnCa2004}.  The concentration of CNO and other relatively light species might show small variations but Fe is not expected to be affected.  \citet{MichaudVaVa83} however suggested that \gr{}s should
lead to overabundances of at least some metals in those stars where settling causes underabundances of \He.  \citet{GlaspeyMiMoetal89}
have confirmed the overabundance of Fe in one star of one cluster but at the limit of detection and this observation required confirmation.
This prediction has now been confirmed in many clusters \citep{BehrCoMcetal99,MoehlerSwLaetal2000,FabbianReGretal2005,PaceRePietal2006} but in particular by \citet{Behr2003} for M15. Overabundances of Fe by factors of 50--100 are seen in nearly all HB stars with $\teff > 11500$\,K while the cooler ones have
the same Fe abundance as cluster's giants (see Fig. 2). 

In stellar evolution calculations, the surface concentrations can be affected by the exterior boundary conditions.   In the calculations of \citet{MichaudRiRi2008}, the simplest assumption was made, that of a mixed outer zone without any mass loss.  
The concentration variations within a 0.61\msol{} HB  star, 30 Myr after ZAHB, are illustrated in the left panel of  Fig.\,\ref{fig:EventailHB}.  
One notes that, in this HB model, all species more massive than O, except S, are overabundant in the external region.  There is a second abundance peak of metals that  is progressively deeper in the star from S (at $\log (\Delta m/\Mstar) \sim -5$) to Fe (at $\log (\Delta m/\Mstar) \sim -3$). This is caused by \gr{}s being linked to electronic shells and heavier elements getting into a given electronic shell at higher $T$.

The density dependence of the turbulent diffusion coefficient  was adjusted to reproduce approximately the observations of Fe in one of the stars observed by \citet{Behr2003} in M15. 
In practice, the outer $(\log (\Delta m/\Mstar) \sim -7$ corresponding to the envelope above $T \sim 10^5$\,K) was mixed.  
The same model reproduced reasonably well the observations in other high \teff{} stars of that cluster as may be seen in Fig.\,\ref{fig:Fe}.  
Furthermore, as may be seen in Fig. 11 and 12 of \citet{MichaudRiRi2008} the other anomalies are also reasonnably well reproduced.
This is a striking confirmation of the role of \gr{}s in HB stars.

Some hints as to possible additional separation mechanisms may be suggested 
by the anomalies which are not as well reproduced.  In particular as may be seen from the right
panel of Fig. \ref{fig:Fe},  He  is expected to be underabundant with the turbulence model determined by the Fe abundance
and indeed it is observed to be underabundant.  But the observed underabundance is larger than expected from the calculations.  If one assumes that the mixed zone only comprises the region above $T \sim 30000$\,K, expected He underabundances are increased by many orders of magnitude.  
Given the size of the observed error bars of the He abundance, it may be premature to conclude 
that observations require separation within the outer $\log (\Delta m/\Mstar) \sim -7$.
 The right panel of Fig.\,\ref{fig:Fe} is however suggestive of the potential role of additional separation in  atmospheric regions.

\section{Pop I, AmFm and HgMn stars}
\label{sec:PopI}
\begin{figure}[t!]
\centerline{
\includegraphics[width=0.5\hsize]{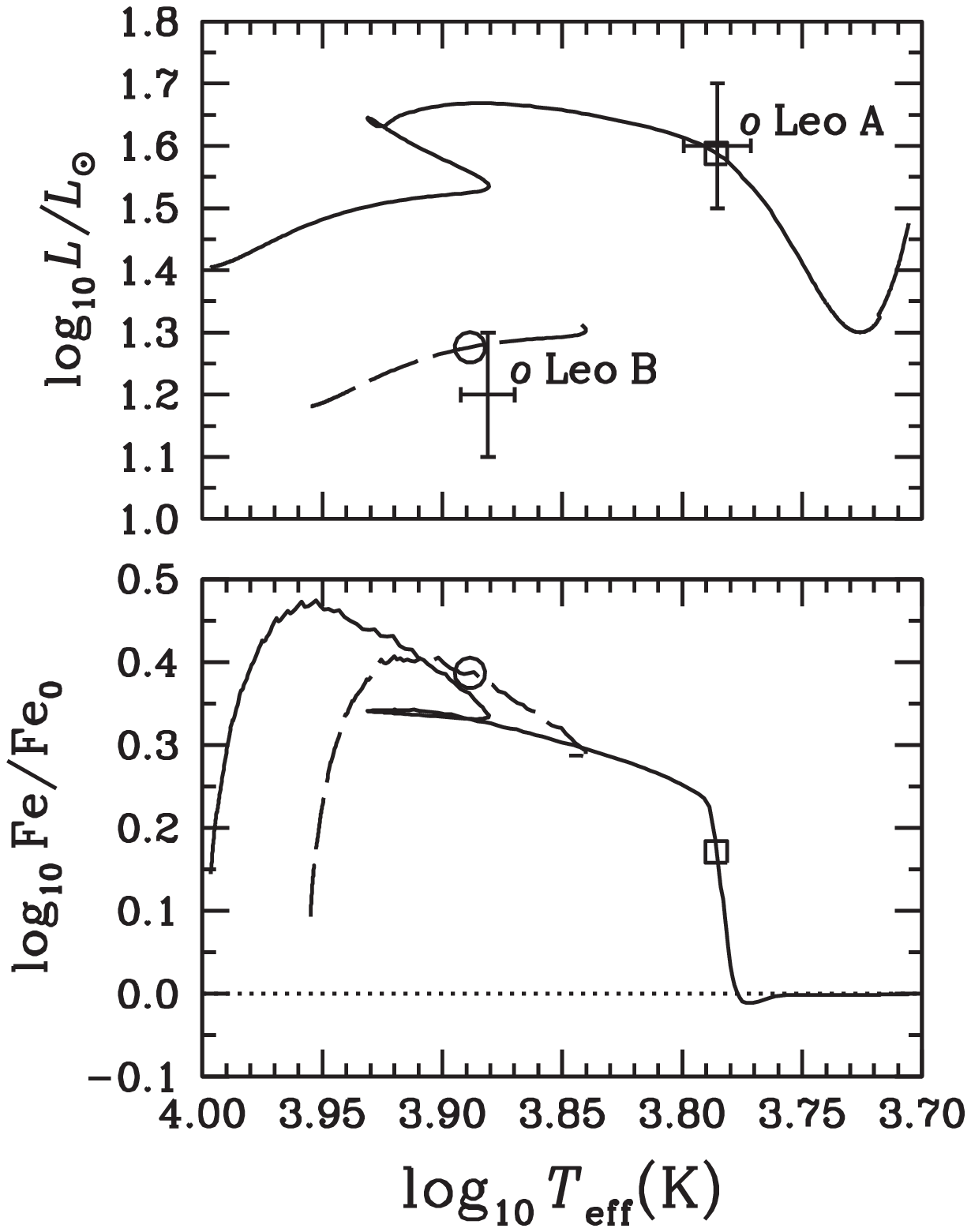}
\includegraphics[width=0.45\hsize]{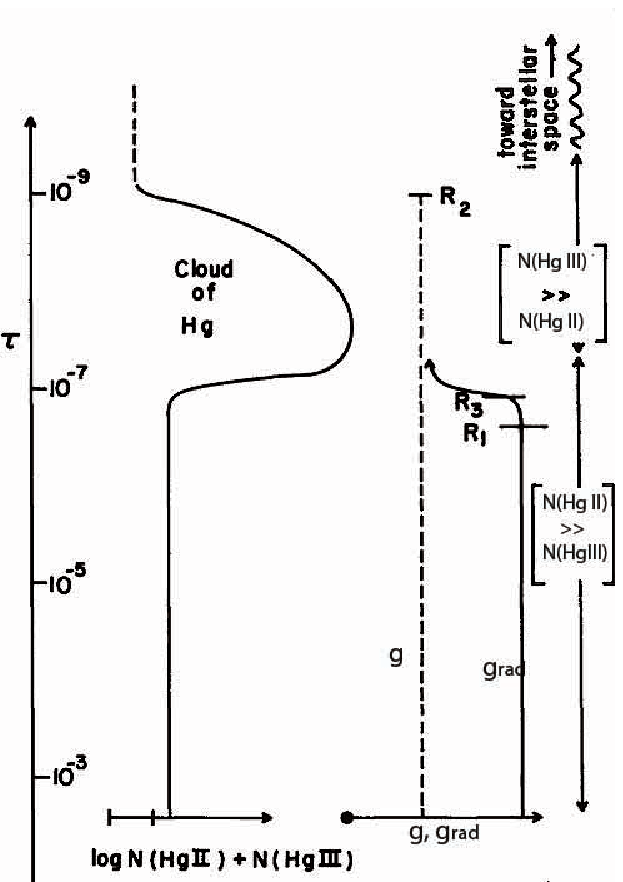}
}
\caption{
\emph{Left panel}, in the upper part are shown, in the HR diagram, the observed positions indicated by crosses of the primary $o{}$\,Leo\,A and the secondary $o$\,Leo\,B as determined by 
 \citet{Griffin2002}.  One model is shown for the A component: 2.24\msol{} (solid line) and one for the B component:
 1.97\msol{} (dashed line).  When the computed primary has the observed
\teff{} (square), the 1.97\msol{} secondary has the \teff{} and $L$ indicated by a circle, which is within the observed error bars.  In the bottom part is shown the surface Fe abundance divided by the original value.
\emph{Right panel},  schematic description of  an upper atmosphere assumed stable enough for separation 
to be important, from \citealt{MichaudReCh74}.  In this example, Hg is pushed from the stellar interior by \gr{}(Hg)
to the point where $\gr \sim g$.  The reduction of the \gr{} as $\tau$ decreases is caused by its progressive ionization into Hg\,III where it has fewer lines in the outgoing radiation flux than Hg\,II.  Mercury accumulates there and forms a cloud. }
\label{fig:Hg}
\end{figure}
The evolution code used above for HB stars had been used before by \citet{RicherMiTu2000} to model AmFm stars.  In their
model, the separation  occurs relatively deep in the star, at  $\log (\Delta m/\Mstar) \sim -5$  where Ca is in the Ne configuration.  The concentration variations within a typical AmFm star may be seen in the right panel of Fig.\,\ref{fig:EventailHB}.  Results of the calculations were compared to observations for a number of stars.  For instance in Fig.\,19 of \citet{RicherMiTu2000} it is shown that there is reasonnable agreement for 10 of the 11 chemical elements common to the observations and calculations for the Hyades star 68  Tau.  In Fig.\,18 of the same paper
reasonnable agreement is found for 12 out of 16 elements for Sirius.

While an alternate  model for AmFm stars, assuming separation just below the H convection zone, is a possible alternative
\citep{Watson71, Smith73,MichaudTaChetal83,Alecian96}, a strong argument for the \citet{RicherMiTu2000} model is 
offered by the observations of the system $o$ Leo whose  primary and secondary are both observed to be AmFm stars.
 The $M$, $L$ and \teff{} are relatively well constrained by the observations. The A component is observed to be unusually cool for an AmFm star. As may be seen in the upper left hand panel of Fig.\,\ref{fig:Hg}, the A component is
 just about to become a red giant and is in a very rapid evolutionary stage.  The presence in a binary system confirms 
 its position in the HR diagram and so the rapid evolutionary stage it is in.   In the lower left hand panel, it is seen that the Fe concentration marks the AmFm character for both components even though the primary is in an advanced subgiant state.  It is just about to loose this Fe overabundance however.  A model with only superficial abundance anomalies on the \MS{} would not maintain them to the evolutionary stage of $o{}$\,Leo\,A.  That the outer 
 $10^{-5}$ of the stellar mass have modified concentrations in the \citet{RicherMiTu2000}
  model explains  that such a subgiant can keep the AmFm characteristics.

In the right hand panel of Fig. [\ref{fig:Hg}] is found a schematic description of a HgMn atmosphere which is expected to have no outer convection zone.  Clouds of various species are expected to form and can  modify the concentration variations that come from the interior. This relatively simple model was found to explain many abundance anomalies observed on HgMn stars.  For instance, \citet{BorsenbergerMiPr79} reproduced reasonably well the \citet{Heacox79}
observations of Sr.  \citet{JomaronDwAl99} measured Mn surface abundances in agreement with the \citet{AlecianMi81} calculations.  

The most careful calculation yet done of \gr{} in HgMn stars is probably  that of \citet{ProffittBrLeetal99}
for Hg.  It included careful evaluation of  momentum sharing between ions.  It was however carried for a homogeneous Hg abundance in the atmosphere.  They obtain that an overabundance by a factor of 10$^4$ of Hg can be supported by \gr{}(Hg) instead of the observed 10$^5$ overabundance.  This is not that bad an agreement taking into account that
it is most likely that a cloud of Hg actually forms in the atmosphere and that this would modify the Hg supported in the 
atmospheric region.  Their assumption of a completely 
mixed outer atmosphere would certainly not explain the Hg isotope anomalies observed on HgMn stars \citep{Preston71}. 
Hg isotope anomalies seem to require separation in the outer atmosphere \citep{MichaudReCh74}. The presence of isotope anomalies is probably the  strongest argument in favor of separation going on in the  atmosphere in addition to the bottom of the mixed zone, both in HgMn stars and the probably related HB stars.

\section{Conclusion}
\label{sec:Conclusion}
Assuming the outer $\sim 10^{-7}$ of the mass of HB stars to be mixed one obtains, with complete evolutionary models from the zero age \MS{}, abundance anomalies corresponding to those observed on M15 (\S \ref{sec:Anomalies}).  
Similar success has been obtained for AmFm stars (\S \ref{sec:PopI}). 
However such an assumption  is almost certainly an oversimplification.  
There are some indications of atmospheric effects in the underabundances of \He{}.  If the values quoted by 
\citet{Behr2003} are not too affected by NLTE effects, larger underabundances of He are observed than expected in our model.  Our model leads to underabundaces of He but not nearly as large as the underabundances claimed by Behr.  However as mentioned in \S \ref{sec:PopI}, there are additional separations going on in the atmospheric regions of HgMn stars to which HB stars are very closely related as originally suggested by \citet{SargentSe67, SargentSe68}.  
The similarity in \teff{} and $\log g$ with HgMn stars suggests that 
separation in the atmosphere  may play a role in HB stars also.  
Note  from the right hand panel of Fig. \ref{fig:Fe} that the model with the thinnest mixed zone develops much smaller He abundances than the one with complete mixing of the outer $\sim 10^{-7}$ of the mass.   Extensive  \gr{} calculations in HB star atmospheres such as those of
\citet{HuibonhoaLeHa2000} would be required in order to couple surface and interior concentration variations more precisely.

The calculations described in \S \ref{sec:Anomalies} explain most of the \teff{} dependence of [Fe/H]
 seen in Fig. \ref{fig:Fe}, but not all of it.
A potential  link between the observed anomalies and rotation in   HB stars has been analyzed in  \citet{QuievyChMietal2007}.  From Figure 1 of that paper, there seems to be little doubt that rotation plays a role;
furthermore according to their calculations involving a parameter free meridional circulation model,  rotation  explains why stars with $\teff < 11000$ K have normal Fe abundances and why one of the hotter stars in M15 (the one indicated by an arrow in the right hand panel of Fig. \ref{fig:Fe}) does not have abundance anomalies.  While atomic diffusion driven by \gr{} 
appears to play the major role in explaining the anomalies, the slow rotation of the hotter HB stars appears to be important also.  It remains to explain why they rotate so slowly.  Similarly it is not understood why the \MS{}
equivalent, the HgMn stars, rotate as slowly as they do.  In this case one may assume a small original rotation rate but it is not clear that one may extend that  assumption to HB stars.  The slow rotation of the HB stars with $\teff > 11\,500$\,K remains unexplained.

Finally mass loss may also play a role. As shown by Vick and Michaud (C{\_}Vick, these proceedings), 
including mass loss in models for AmFm stars 
leads to approximately as good agreement with observations as obtained in models assuming  turbulent mixing.  
Observational tests are suggested to distinguish between the two processes but current observations are not accurate enough to allow the elimination of  one or the other model. It would be interesting  to determine if mass loss could be as succesful as turbulent mixing for HB stars also.

\acknowledgements
This research was partially supported at  the Universit\'e de Montr\'eal 
by NSERC. We thank the R\'eseau qu\'eb\'ecois de calcul de haute
performance (RQCHP)
for providing us with the computational resources required for this
work. 
\newcommand{\newblock}{}


\begin{thebibliography}{43}
\expandafter\ifx\csname natexlab\endcsname\relax\def\natexlab#1{#1}\fi

\bibitem[{Alecian(1996)}]{Alecian96}
Alecian, G.: 1996, \textit{Astron. Astrophys.}, \textbf{310}, 872

\bibitem[{Alecian \& Michaud(1981)}]{AlecianMi81}
Alecian, G. \& Michaud, G.: 1981, \textit{Astrophys. J.}, \textbf{245}, 226

\bibitem[{{Behr}(2003)}]{Behr2003}
{Behr}, B.~B.: 2003, \textit{Astrophys. J., Suppl. Ser.}, \textbf{149}, 67

\bibitem[{Behr {et~al.}(1999)Behr, Cohen, McCarthy, \&
  Djorgovski}]{BehrCoMcetal99}
Behr, B.~B., Cohen, J.~G., McCarthy, J.~K., \& Djorgovski, S.~G.: 1999,
  \textit{Astrophys. J., Lett.}, \textbf{517}, L135

\bibitem[{Borsenberger {et~al.}(1979)Borsenberger, Michaud, \&
  Praderie}]{BorsenbergerMiPr79}
Borsenberger, J., Michaud, G., \& Praderie, F.: 1979, \textit{Astron. Astrophys.}, \textbf{76},
  287

\bibitem[{Burgers(1969)}]{Burgers69}
Burgers, J.~M.: 1969, \textit{Flow Equations for Composite Gases} (New York: Academic
  Press)

\bibitem[{{Fabbian} {et~al.}(2005){Fabbian}, {Recio-Blanco}, {Gratton}, \&
  {Piotto}}]{FabbianReGretal2005}
{Fabbian}, D., {Recio-Blanco}, A., {Gratton}, R.~G., \& {Piotto}, G.: 2005,
  \textit{Astron. Astrophys.}, \textbf{434}, 235

\bibitem[{Glaspey {et~al.}(1989)Glaspey, Michaud, Moffat, \&
  Demers}]{GlaspeyMiMoetal89}
Glaspey, J.~W., Michaud, G., Moffat, A.~F.~J., \& Demers, S.: 1989, \textit{Astrophys.
  J.}, \textbf{339}, 926

\bibitem[{{Gratton} {et~al.}(2004){Gratton}, {Sneden}, \&
  {Carretta}}]{GrattonSnCa2004}
{Gratton}, R., {Sneden}, C., \& {Carretta}, E.: 2004, \textit{Ann. Rev. Astron.
  Astrophys.}, \textbf{42}, 385

\bibitem[{{Griffin}(2002)}]{Griffin2002}
{Griffin}, R.~E.: 2002, \textit{Astron. J.}, \textbf{123}, 988

\bibitem[{{Heacox}(1979)}]{Heacox79}
{Heacox}, W.~D.: 1979, \textit{Astrophys. J., Suppl. Ser.}, \textbf{41}, 675

\bibitem[{{Hui-Bon-Hoa} {et~al.}(2000){Hui-Bon-Hoa}, {LeBlanc}, \&
  {Hauschildt}}]{HuibonhoaLeHa2000}
{Hui-Bon-Hoa}, A., {LeBlanc}, F., \& {Hauschildt}, P.~H.: 2000, \textit{Astrophys. J.,
  Lett.}, \textbf{535}, L43

\bibitem[{{Jomaron} {et~al.}(1999){Jomaron}, {Dworetsky}, \&
  {Allen}}]{JomaronDwAl99}
{Jomaron}, C.~M., {Dworetsky}, M.~M., \& {Allen}, C.~S.: 1999, \textit{Mon. Not. R.
  Astron. Soc.}, \textbf{303}, 555

\bibitem[{King {et~al.}(1998)King, Stephens, Boesgaard, \&
  Deliyannis}]{KingStBoetal98}
King, J.~R., Stephens, A., Boesgaard, A.~M., \& Deliyannis, C.~F.: 1998, \textit{Astron.
  J.}, \textbf{115}, 666

\bibitem[{Michaud(1981)}]{Michaud81}
Michaud, G.: 1981, in \textit{Chemically Peculiar Stars of the Upper Main Sequence}
  (Li\`ege: Universit\'e de Li\`ege), 355--363

\bibitem[{{Michaud} {et~al.}(1974){Michaud}, {Reeves}, \&
  {Charland}}]{MichaudReCh74}
{Michaud}, G., {Reeves}, H., \& {Charland}, Y.: 1974, \textit{Astron. Astrophys.}, \textbf{37},
  313

\bibitem[{{Michaud} {et~al.}(2005){Michaud}, {Richer}, \&
  {Richard}}]{MichaudRiRi2005}
{Michaud}, G., {Richer}, J., \& {Richard}, O.: 2005, \textit{Astrophys. J.}, \textbf{623}, 442

\bibitem[{Michaud {et~al.}(2007)Michaud, Richer, \& Richard}]{MichaudRiRi2007}
Michaud, G., Richer, J., \& Richard, O.: 2007, \textit{Astrophys. J.}, \textbf{670}, 1178

\bibitem[{Michaud {et~al.}(2008)Michaud, Richer, \& Richard}]{MichaudRiRi2008}
Michaud, G., Richer, J., \& Richard, O.: 2008, \textit{Astrophys. J.}, submitted

\bibitem[{Michaud {et~al.}(1983{\natexlab{a}})Michaud, Tarasick, Charland, \&
  Pelletier}]{MichaudTaChetal83}
Michaud, G., Tarasick, D., Charland, Y., \& Pelletier, C.: 1983{\natexlab{a}},
  \textit{Astrophys. J.}, \textbf{269}, 239

\bibitem[{Michaud {et~al.}(1983{\natexlab{b}})Michaud, Vauclair, \&
  Vauclair}]{MichaudVaVa83}
Michaud, G., Vauclair, G., \& Vauclair, S.: 1983{\natexlab{b}}, \textit{Astrophys. J.},
  \textbf{267}, 256

\bibitem[{Moehler {et~al.}(2000)Moehler, Sweigart, Landsman, \&
  Heber}]{MoehlerSwLaetal2000}
Moehler, S., Sweigart, A.~V., Landsman, W.~B., \& Heber, U.: 2000, \textit{Astron.
  Astrophys.}, \textbf{360}, 120

\bibitem[{{Pace} {et~al.}(2006){Pace}, {Recio-Blanco}, {Piotto}, \&
  {Momany}}]{PaceRePietal2006}
{Pace}, G., {Recio-Blanco}, A., {Piotto}, G., \& {Momany}, Y.: 2006, \textit{Astron.
  Astrophys.}, \textbf{452}, 493

\bibitem[{{Paczy{\'n}ski}(1970)}]{Paczynski70}
{Paczy{\'n}ski}, B.: 1970, \textit{Acta~Astron.}, \textbf{20}, 195

\bibitem[{{Preston}(1971)}]{Preston71}
{Preston}, G.~W.: 1971, \textit{Astrophys. J., Lett.}, \textbf{164}, L41

\bibitem[{Proffitt(1994)}]{Proffitt94}
Proffitt, C.~R.: 1994, \textit{Astrophys. J.}, \textbf{425}, 849

\bibitem[{{Proffitt} {et~al.}(1999){Proffitt}, {Brage}, {Leckrone}, {Wahlgren},
  {Brandt}, {Sansonetti}, {Reader}, \& {Johansson}}]{ProffittBrLeetal99}
{Proffitt}, C.~R., {Brage}, T., {Leckrone}, D.~S., {et~al.}: 1999, \textit{Astrophys.
  J.}, \textbf{512}, 942

\bibitem[{Proffitt \& Michaud(1991)}]{ProffittMi91}
Proffitt, C.~R. \& Michaud, G.: 1991, \textit{Astrophys. J.}, \textbf{371}, 584

\bibitem[{Quievy {et~al.}(2007)Quievy, Charbonneau, Michaud, \&
  Richer}]{QuievyChMietal2007}
Quievy, D., Charbonneau, P., Michaud, G., \& Richer, J.: 2007, \textit{Astrophys. J.},
  submitted, 00

\bibitem[{Richard {et~al.}(2002{\natexlab{a}})Richard, Michaud, \&
  Richer}]{RichardMiRi2002}
Richard, O., Michaud, G., \& Richer, J.: 2002{\natexlab{a}}, \textit{Astrophys. J.}, \textbf{580},
  1100

\bibitem[{Richard {et~al.}(2005)Richard, Michaud, \& Richer}]{RichardMiRi2005}
Richard, O., Michaud, G., \& Richer, J.: 2005, \textit{Astrophys. J.}, \textbf{619}, 538

\bibitem[{Richard {et~al.}(2002{\natexlab{b}})Richard, Michaud, Richer,
  Turcotte, Turck-Chieze, \& VandenBerg}]{RichardMiRietal2002}
Richard, O., Michaud, G., Richer, J., {et~al.}: 2002{\natexlab{b}}, \textit{Astrophys.
  J.}, \textbf{568}, 979

\bibitem[{Richer {et~al.}(2000)Richer, Michaud, \& Turcotte}]{RicherMiTu2000}
Richer, J., Michaud, G., \& Turcotte, S.: 2000, \textit{Astrophys. J.}, \textbf{529}, 338

\bibitem[{Rogers \& Iglesias(1992)}]{RogersIg92a}
Rogers, F.~J. \& Iglesias, C.~A.: 1992, \textit{Astrophys. J.}, \textbf{401}, 361

\bibitem[{Sargent \& Searle(1967)}]{SargentSe67}
Sargent, W.~L.~W. \& Searle, L.: 1967, \textit{Astrophys. J., Lett.}, \textbf{150}, L33

\bibitem[{Sargent \& Searle(1968)}]{SargentSe68}
Sargent, W.~L.~W. \& Searle, L.: 1968, \textit{Astrophys. J.}, \textbf{152}, 443

\bibitem[{Smith(1973)}]{Smith73}
Smith, M.~A.: 1973, \textit{Astrophys. J., Suppl. Ser.}, \textbf{25}, 277

\bibitem[{{Sweigart}(1987)}]{Sweigart87}
{Sweigart}, A.~V.: 1987, \textit{Astrophys. J., Suppl. Ser.}, \textbf{65}, 95

\bibitem[{{Sweigart}(1994)}]{Sweigart94}
{Sweigart}, A.~V.: 1994, in \textit{Hot Stars in the Galactic Halo}, Cambridge University
  Press, ed. S.~J. {Adelman}, A.~R. {Upgren}, \& C.~J. {Adelman}, 17

\bibitem[{Turcotte {et~al.}(1998)Turcotte, Richer, Michaud, Iglesias, \&
  Rogers}]{TurcotteRiMietal98}
Turcotte, S., Richer, J., Michaud, G., Iglesias, C., \& Rogers, F.: 1998,
  \textit{Astrophys. J.}, \textbf{504}, 539

\bibitem[{{VandenBerg}(1985)}]{VandenBerg85}
{VandenBerg}, D.~A.: 1985, \textit{Astrophys. J., Suppl. Ser.}, \textbf{58}, 711

\bibitem[{VandenBerg {et~al.}(2002)VandenBerg, Richard, Michaud, \&
  Richer}]{VandenBergRiMietal2002}
VandenBerg, D.~A., Richard, O., Michaud, G., \& Richer, J.: 2002, \textit{Astrophys. J.},
  \textbf{571}, 487

\bibitem[{Watson(1971)}]{Watson71}
Watson, W.~D.: 1971, \textit{Astron. Astrophys.}, \textbf{13}, 263

%
%
%
%
%
%
\end{thebibliography}

\end{document}